\documentclass[preprint]{revtex4-1}
\usepackage{amsmath,amssymb}
\usepackage{graphicx}

\newcommand\Shat{\mathbf{\hat{S}}}
\newcommand\Lhat{\mathbf{\hat{L}}}
\newcommand\J{\mathbf{J}}

\newcommand\EE{\mathbf{E}}
\newcommand\HH{\mathbf{B}}

\newcommand\p{\mathbf{p}}
\newcommand\rr{\mathbf{r}}
\newcommand\PP{\mathbf{P}}

\newcommand\xhat{\mathbf{\hat{x}}}
\newcommand\uhat{\mathbf{\hat{u}}}
\newcommand\yhat{\mathbf{\hat{y}}}

\newcommand\Field{|\Psi_{\omega,\nu}\rangle}
\newcommand\Gen{\left(-i\beta_x\Lambda P_x/|\PP|\right)}
\newcommand\Gentwo{\left(-i\beta_x P_x/|\PP|\right)}
\newcommand\Genthree{\left(-i(\nu\beta_x/\omega) P_x\right)}
\newcommand\Longint{\int_0^{\infty} d\omega\int_0^{\pi}d\theta\sin\theta\int_{-\pi}^{\pi}d\phi}

\begin{document}
\title[Transformations generated by the spin and orbital angular momentum operators]{On the transformations generated by the electromagnetic spin and orbital angular momentum operators.}
\author{Ivan Fernandez-Corbaton$^{1,2}$, Xavier Zambrana-Puyalto$^{1,2}$ and Gabriel Molina-Terriza$^{1,2}$}
\address{$^1$ Department of Physics \& Astronomy, Macquarie University, NSW 2109, Australia\\$^2$ ARC Center for Engineered Quantum Systems, NSW 2109, Australia}
\email{ivan.fernandez-corbaton@mq.edu.au}
\begin{abstract}
We present a study of the properties of the transversal ``spin angular momentum'' and ``orbital angular momentum'' operators. We show that the ``spin angular momentum'' operators are generators of spatial translations which depend on helicity and frequency and that the ``orbital angular momentum'' operators generate transformations which are a sequence of this kind of translations and rotations. We give some examples of the use of these operators in light matter interaction problems. Their relationship with the helicity operator allows to involve the electromagnetic duality symmetry in the analysis. We also find that simultaneous eigenstates of the three ``spin'' operators and parity define a type of standing modes which has been recently singled out for the interaction of light with chiral molecules. With respect to the relationship between ``spin angular momentum'', polarization, and total angular momentum, we show that, except for the case of a single plane wave, the total angular momentum of the field is decoupled from its vectorial degrees of freedom even in the regime where the paraxial approximation holds. Finally, we point out a relationship between the three ``spin'' operators and the spatial part of the Pauli-Lubanski four vector.
\end{abstract}
\maketitle
\noindent The total angular momentum of an electromagnetic field can be split into gauge independent two parts \cite[chap. XXI \S23]{Messiah1958}, \cite[probl. 7.27]{Jackson1998}, \cite{VanEnk1994},\cite[chap. 10.6.2]{Mandel1995}.  These two parts are commonly referred to as spin and orbital ``angular momenta''. Even though these names have stuck in the literature, it is well known that as operators, neither of them obey the commutation relations that define angular momenta \cite{VanEnk1994}. After the seminal work of Van Enk and Nienhuis in \cite{VanEnk1994}, several authors have studied the properties of these operators \cite{Jauregui2005,Barnett2010,Bliokh2010}. 

Since neither the ``spin angular momentum'' $\Shat$ nor the ``orbital angular momentum'' $\Lhat$ are angular momenta, they are not the generators of rotations. Identifying the transformation generated by a given operator makes such operator useful for the study of light matter interactions from the point of view of symmetries and conservation laws \cite[chap. XIV]{Messiah1958}. For example, due to spherical symmetry we know that the resonant electromagnetic modes of a sphere have to be eigenstates of angular momentum \cite[chap. 10]{Jackson1998}.

In this article, we study these two vector operators, $\Shat$ and $\Lhat$. The rest of the article is organized as follows. In Sec. \ref{sec:setting} we specify the mathematical setting and notation that we will use in the article and touch upon the helicity operator and its associated electromagnetic duality symmetry transformation. In Sec. \ref{sec:introduction} we give a short introduction to the current theory of the $\Shat$ and $\Lhat$ operators. In Sec. \ref{sec:using}, and with the help of the helicity operator, we to obtain the definitions of $\Shat$ and $\Lhat$ as abstract operators. We also show that those definitions are consistent with the existing theory. Section \ref{sec:transformations} contains the derivation of the transformations generated by $\Shat$ and $\Lhat$, and Sec. \ref{sec:examples} examples of their uses in interactions of the electromagnetic field with material systems. In Sec. \ref{sec:polarization} we address the related question of the relationship between the polarization degrees of freedom and angular momentum and prove that angular momentum and the polarization degrees of freedom are decoupled except in the case of a single plane wave. To finalize the analysis, in Sec. \ref{sec:shat}, we point out a connection between the ``spin angular momentum'' vector of operators and the spatial part of the Pauli-Lubanski four-vector used in relativistic field theory.

\section{Mathematical setting, notation and the helicity operator}\label{sec:setting}
We will make use of the formal setting and tools of Hilbert spaces. Our Hilbert space will be that of the transverse solutions of Maxwell's equations. We will denote this space by $\mathbb{M}$, and the vectors in it using the ``ket'' notation $|\Psi\rangle$. The vectors in $\mathbb{M}$ are acted upon by linear operators that map them back to $\mathbb{M}$. Transformations like rotations, space translations and time translations, are represented by unitary operators in $\mathbb{M}$. The generators of those transformations are also operators in $\mathbb{M}$, i.e, angular momentum $\J$, momentum $\PP$ and $P_0$ for the aforementioned transformations, respectively. To indicate that a vector is an eigenstate of an operator, we will write its eigenvalue in the subscript. For example $|\Psi_{\p}\rangle$ denotes an eigenstate of the three components of the linear momentum operator $\PP$. When we write that a vector has a well defined value of operator X equal to x, we mean that it is an eigenstate of X with eigenvalue x. In order to avoid confusion with other uses of the term {\em vector}, we will use the term {\em mode} instead of {\em vector} to refer to members of $\mathbb{M}$. We will use {\em frequency} ($\omega$) to denote the eigenvalues of the generator of time translations $P_0$, and, without loss of generality, use only positive frequencies $(\omega>0)$.

We use this formal setting because the arguments and derivations will be based on commutation relations between operators and the transformation properties of modes in $\mathbb{M}$. These kind of reasoning only relies on the underlying algebraic structure of $\mathbb{M}$, and is therefore independent of the particular representation of $\mathbb{M}$. For example: Rotations along any axis $\uhat$ commute with time translations. This statement applies to classical fields in the common ``coordinate'' representation of $\mathbb{M}$ ($\EE(\rr,t),\HH(\rr,t)$), and also to their momentum space representation \cite[Chap. I.B]{Cohen1997} $(\mathcal{E(\p)},\mathcal{B(\p)})$. It also applies, for example, to single photon states in quantized electromagnetism. As a consequence, we can construct in both representations simultaneous eigenstates of rotations and time translations or, equivalently, of their generators $\uhat\cdot\J$ and $P_0$. 

Helicity, the generator of the electromagnetic duality transformation \cite{Calkin1965,Zwanziger1968}, will play an important role in the discussion. Helicity is defined as the projection of the angular momentum vector operator $\J$ onto the direction of the linear momentum \cite[eq. 8.4-5]{Tung1985}:
\begin{equation}
	\label{eq:hel}
	\Lambda=\frac{\J\cdot\PP}{|\PP|}.
\end{equation}
Tools for the use of helicity and duality in light matter interactions are already available \cite{FerCor2012b,FerCor2012p,Zambrana2013b,FerCor2013}, including the conditions that a scatterer must meet in order for it to be invariant under duality transformations, i.e. to preserve the helicity of the field upon scattering. In the momentum (or plane wave) representation of classical fields, there is an intuitive operational definition of helicity: An electromagnetic field has a well defined helicity only when all the plane waves in its decomposition have the same polarization handedness with respect to their corresponding momentum vectors. For classical electromagnetic fields and single photon Fock states, helicity, like handedness, can take two values. That is, $\Lambda$ has two eigenvalues of opposite sign $\nu=\pm1$. We will restrict the derivations to this case. 

\section{Introduction to $\Shat$ and $\Lhat$}\label{sec:introduction}
We start the discussion from the expression of the ``spin angular momentum'' operators $\Shat$ in two different representations. The first one is the Fock space of quantized modes with well defined momentum ($\p$) and helicity ($\pm$). It can be found for instance in \cite[chap XXI. prob. 7]{Messiah1958},\cite[chap. 10.6.3]{Mandel1995} and \cite{VanEnk1994}:
\begin{equation}
	\label{eq:second}
	\Shat_{F}=\int{d\p}\left(\hat{n}_{\p,+}-\hat{n}_{\p,-}\right)\frac{\p}{|\p|},
\end{equation}
where the $\hat{n}_{\p,\pm}$ are the number operators. 

The second one is the momentum representation of classical fields. The action of $\Shat$ on a single mode of well defined momentum $\p$ and arbitrary polarization $\tau$, $\mathcal{F_{\tau}(\p)=\hat{\tau} \exp(\p\cdot\rr)}$, is written in \cite[eq. 6]{Bliokh2010} to be:
\begin{equation}
	\label{eq:first}
	\Shat_{m}\mathcal{F_{\tau}(\p)}=\frac{\p}{|\p|}\left(\frac{\sum_{i=1}^3 p_iS_i}{|\p|}\right)\mathcal{F_{\tau}(\p)}=\frac{\p}{|\p|}\left(\frac{\p}{|\p|}\cdot \mathbf{S}\right)\mathcal{F_{\tau}(\p)},
\end{equation}
where the components of $\mathbf{S}$ are the three spin one matrices \cite[chap. XIII \S 21]{Messiah1958}. In both (\ref{eq:second}) and (\ref{eq:first}), $\p$ are numbers: The three momenta eigenvalues of the modes on which the operators act on. 

Let us now consider states with well defined helicity $\nu=\pm1$: $\mathcal{F_{\nu}(\p)}$. After noting that $\J\cdot\PP=\left(\rr\times\mathbf{L}+\mathbf{S}\right)\cdot\PP=\mathbf{S}\cdot\PP$, so that
\begin{equation}
	\Lambda=\frac{\J\cdot\PP}{|\PP|}=\frac{\mathbf{S}\cdot\PP}{|\PP|},
\end{equation}
it follows from (\ref{eq:first}) that
\begin{equation}
\label{eq:au}
	\Shat_{m}\mathcal{F_{\nu}(\p)}=\frac{\p\nu}{|\p|}\mathcal{F_{\nu}(\p)}.
\end{equation}
From (\ref{eq:au}), the expression of $\Shat_{m}$ using abstract modes in $\mathbb{M}$ can be deduced to be:
\begin{equation}
	\label{eq:mom}
	\Shat_{m}\equiv\sum_{\nu}\int{d\p}\frac{\p\nu}{|\p|}|\Psi_{\p\nu}\rangle\langle \Psi_{\p\nu}|=\int{d\p}\frac{\p}{|\p|}\left(|\Psi_{\p+}\rangle\langle \Psi_{\p+}|-|\Psi_{\p-}\rangle\langle \Psi_{\p-}|\right).
\end{equation}

One important remark is that $\mathbf{S}$ enters the definition of $\Shat_{m}$ in the dot product $(\p/|\p|)\cdot \mathbf{S}$. The components of $\mathbf{S}$ by themselves are not operators in $\mathbb{M}$ because they break the transversality of the field \cite{VanEnk1994,Cohen1997,Barnett2010}. So do the components of $\mathbf{L}=\J-\mathbf{S}$. While it is true that $\J=\mathbf{S}+\mathbf{L}=\Shat+\Lhat$, there is a fundamental difference between the two pairs of vector operators. In general, the action of the components of $\mathbf{L}$ and $\mathbf{S}$ break the transversality condition and take a mode in $\mathbb{M}$ outside of $\mathbb{M}$. On the other hand $\Shat$ and $\Lhat$ map any mode in $\mathbb{M}$ back onto $\mathbb{M}$. We will use the carets in $\Shat$ and $\Lhat$ to distinguish these two transverse operators from the non-transverse ones.

The commutation relations between $\Shat$ and $\Lhat=\J-\Shat$ were found to be exactly the same in the two representations of Eqs. (\ref{eq:second}) and (\ref{eq:first}) (\cite{VanEnk1994b,Bliokh2010}), reflecting the fact that they represent the same algebraic structure in $\mathbb{M}$. With $\varepsilon_{jkl}$ denoting the totally antisymmetric tensor with $\varepsilon_{123}=1$, the commutation relations read
\begin{equation}
	\label{eq:commu}
	\begin{split}
		[\hat{S}_j,\hat{S}_k]=0&,\ [\hat{L}_j,\hat{L}_k]=i\sum_l\varepsilon_{jkl}(\hat{L}_l-\hat{S}_l),\\
		  [\hat{S}_j,\hat{L}_k]&=i\sum_l\varepsilon_{jkl}\hat{S}_l.
	\end{split}
\end{equation}

These are different from the commutation relations that define angular momentum operators  \cite[eq. XIII.3]{Messiah1958}, \cite[chap. 6]{Cohen1991},\cite[chap. 3.1]{Sakurai1993}: $[J_j,J_k]=i\sum_l\varepsilon_{jkl}J_l$. Clearly, neither $\Shat$ nor $\Lhat$ are angular momenta. They do not generate rotations and, consequently, their eigenstates are not necessarily preserved upon interaction with rotationally symmetric systems. On the other hand, they may be preserved by systems without rotational symmetry. We will later give examples of both these cases.

\section{Using helicity to define $\Shat$ and $\Lhat$}\label{sec:using}
From Eqs. (\ref{eq:second}) and (\ref{eq:mom}), it is clear that the representation independent form of $\Shat$ is:
\begin{equation}
	\label{eq:shat}
	\Shat=\Lambda\frac{\PP}{|\PP|}.
\end{equation}
One way to verify this statement is to compute the matrix elements of $\Shat$ between the $|\Psi_{\p\nu}\rangle$ states, which can be scaled to form a complete orthonormal basis of $\mathbb{M}$:
\begin{equation}
	\label{eq:mel}
	\langle \Psi_{\bar{\p}\bar{\nu}}|\Lambda\frac{\PP}{|\PP|}|\Psi_{\p\nu}\rangle=\frac{\p\nu}{|\p|}\langle \Psi_{\bar{\p}\bar{\nu}}|\Psi_{\p\nu}\rangle=\frac{\p\nu}{|\p|}\delta_{\bar{\p}\p}^{\bar{\nu}\nu},
\end{equation}
where the orthonormality of the $|\Psi_{\p\nu}\rangle$ states is expressed by the function $\delta_{\bar{\p}\p}^{\bar{\nu}\nu}$, which is zero unless $\bar{\p}=\p$ and $\bar{\nu}=\nu$. The matrix elements in (\ref{eq:mel}) coincide with those in (\ref{eq:mom}), verifying that $\Shat_m$ is the momentum space representation of $\Shat$. It can also be verified that they also coincide with the corresponding calculations in Fock space.

The definition in (\ref{eq:shat}) implies that $\Lhat=\J-\Lambda\PP/|\PP|$. With these definitions, $\Lhat$ and $\Shat$ meet the commutation relations in (\ref{eq:commu}). This can be verified using that $\Lambda$ commutes with both $\J$ and $\PP$ (\cite[Chaps. 8.4.1, 9.6]{Tung1985}). 


Expression (\ref{eq:shat}) agrees with the interpretation of the transformations generated by the operator in (\ref{eq:second}) given in \cite[sec. 3.3]{VanEnk1994}, where $\Shat_{F}$ is described as the generator of transformations of the momentum space polarization that preserve the transversality of the field. Note that momentum space polarization and coordinate space polarization are two different concepts. The transformation $\exp(-i\boldsymbol\beta\cdot\Shat_{F})$, where $\boldsymbol\beta$ is a real vector, is found in \cite{VanEnk1994} to be a rotation of the polarization of each Fock state of defined momentum by an angle $\theta_{\p}$ which depends on the momentum of the state $\theta_{\p}=\boldsymbol\beta\cdot\p/|\p|$. In order to show the consistency of (\ref{eq:shat}) with the described transformation, we consider the action of $\exp(-i\boldsymbol\beta\cdot\Lambda\PP/|\PP|)$ on modes of defined momentum $|\Psi_{\p}\rangle$:
\begin{equation}
	\label{eq:v}
	\begin{split}
	&\exp(-i\boldsymbol\beta\cdot\Lambda\PP/|\PP|)|\Psi_{\p}\rangle=\\
	&\exp(-i(\boldsymbol\beta\cdot\p/|\p|)\Lambda)|\Psi_{\p}\rangle=\\
	&D(\boldsymbol\beta\cdot\p/|\p|)|\Psi_{\p}\rangle=D(\theta_{\p})|\Psi_{\p}\rangle,
	\end{split}
\end{equation}
where we have used the fact that $|\Psi_{\p}\rangle$ is an eigenstate of $\PP$ and identified $\exp(-i\alpha\Lambda)=D(\alpha)$, the duality transformation, whose action on modes of well defined helicity is:
\begin{equation}
	D(\alpha)|\Psi_{\pm}\rangle=\exp(\mp i\alpha)|\Psi_{\pm}\rangle.
\end{equation}

The interpretation of (\ref{eq:shat}) as the generator of momentum dependent linear polarization rotation is recovered from (\ref{eq:v}) because the plane wave states of linear polarization (TE/TM) are sums and subtractions of helicity eigenstates $|\Psi_{\p,te/tm}\rangle=1/\sqrt{2}(|\Psi_{\p,+}\rangle\pm|\Psi_{\p,-}\rangle)$ \cite[app. B]{FerCor2012b}, and:
\begin{equation}
	\begin{split}
		&D(\theta_{\p})|\Psi_{\p,te/tm}\rangle=\\
		&\frac{1}{\sqrt{2}}\left(\exp(-i\theta_{\p})|\Psi_{\p,+}\rangle\pm\exp(i\theta_{\p})|\Psi_{\p,-}\rangle\right),
	\end{split}
\end{equation}
is a rotation of the linear polarization of both $|\Psi_{\p,te/tm}\rangle$ modes by the same angle $\theta_{\p}=\boldsymbol\beta\cdot\p/|\p|$ found in \cite[sec. 3.3]{VanEnk1994}. Note that, implicitly, the interpretation in \cite[sec. 3.3]{VanEnk1994} also assumes that the eigenvalues of helicity are $\pm 1$.

Another interpretation, this time in the $(\EE(\rr,t),\HH(\rr,t))$ representation, can be found in \cite{Barnett2010,Cameron2012}. In the limit of $|\boldsymbol\beta|\rightarrow 0$, $\boldsymbol\beta\cdot\Shat$ and $\boldsymbol\beta\cdot\Lhat$ are found to generate approximated infinitesimal rotations of the fields around $\boldsymbol\beta$ while leaving either the spatial distribution or the field vector directions unchanged, respectively.

\section{Transformations generated by $\Shat$ and $\Lhat$}\label{sec:transformations}
In order to further understand $\Shat$ and $\Lhat$, and to be able to use them in the study of light matter interactions by means of symmetries and conservation laws, we wish to obtain more insight on the exact action of the transformation $\exp(-i\boldsymbol\beta\cdot\Lambda\PP/|\PP|)$. We will try its action on simultaneous eigenstates of $\Lambda$ and $|\PP|$. Our test modes are hence monochromatic modes with well defined helicity $\nu$ and frequency $\omega=|\p|$ (in units of $c=1$, which we adopt from now on), which we denote by $\Field$. For simplicity, we first study a single component of $\Shat$. We take the first component of $\Shat$, use it to generate the corresponding continuous transformation with a real scalar parameter $\beta_x$ and apply such transformation to $\Field$. We manipulate such expression using the Taylor expansion of the exponential, the fact that helicity and momentum commute, that $\Lambda^2=I$ for transverse electromagnetic fields \cite{FerCor2012p}, and then substitute the operators $\Lambda$ and $|\PP|$ by their eigenvalues $\nu$ and $\omega$:
{\small
\begin{eqnarray}\nonumber
\label{eq:deriv}
&&\exp\Gen\Field=\\\nonumber
&&\sum_{k=0}^\infty \left[\frac{\Gen^{2k}}{(2k)!}+\frac{\Gen^{2k+1}}{(2k+1)!}\right]\Field=\\\nonumber
&&\sum_{k=0}^\infty\left[\frac{\Gentwo^{2k}}{(2k)!}+\frac{\Gentwo^{2k+1}\Lambda}{(2k+1)!}\right]\Lambda^{2k}\Field=\\\nonumber
&&\sum_{k=0}^\infty\frac{\Gentwo^{2k}}{(2k)!}\Field+\nu\sum_{k=1}^\infty \frac{\Gentwo^{2k+1}}{(2k+1)!}\Field=\\
&&\exp\Genthree\Field.
\end{eqnarray}
}
The final expression in (\ref{eq:deriv}) is a translation along the $x$ axis with displacement $\nu\beta_x/\omega$. For a fixed value of $\beta_x$, the magnitude of the translation depends on the frequency of the field. The direction of the translation, i.e. whether it is towards larger or smaller $x$ values, depends on the helicity of the field. This last expression also shows that $\beta_x$ measures the displacement in $2\pi$-units of the wavelength. For example, for $\beta_x=2\pi$ the transformation results in a translation of $\nu\lambda$, i.e. exactly one wavelength.

Since the components of $\Shat$ commute among themselves, the derivation in (\ref{eq:deriv}) can be easily extended to cover the case of $\exp(-i\boldsymbol\beta\cdot\Lambda\PP/|\PP|)$, resulting in:
\begin{equation}
	\label{eq:trans}
	\exp(-i\boldsymbol\beta\cdot\Lambda\PP/|\PP|)\Field=\exp(-i(\nu/\omega)\boldsymbol\beta\cdot\PP)\Field.
\end{equation}
Equation (\ref{eq:trans}) is a translation along the direction of the unitary vector $\hat{\boldsymbol\beta}=\boldsymbol\beta/|\boldsymbol\beta|$ by a displacement equal to $\nu|\boldsymbol\beta|/\omega$. The value of helicity $\nu$ controls the direction of the translation along the $\hat{\boldsymbol\beta}$ axis, and $|\boldsymbol\beta|/\omega$ its absolute value. We conclude that $\Shat$ is the generator of helicity and frequency dependent spatial translations.

The particular case of monochromatic fields in the coordinate representation is worth examining because the action of $\exp(-i\boldsymbol\beta \cdot\Lambda\PP/|\PP|)$ as a helicity dependent translation can be seen very clearly. For a monochromatic field of well defined helicity $\mathbf{F}^{\omega}_{\pm}(x,y,z,t)=\mathbf{\widehat{F}}_{\pm}(x,y,z)\exp(-i\omega t)$, it follows from (\ref{eq:trans}), that the action of $\exp(-i\boldsymbol\beta \cdot\Lambda\PP/|\PP|)$ is
\begin{equation}
\label{eq:F}
\begin{split}
	&\exp(-i\boldsymbol\beta\cdot\Lambda\PP/|\PP|)\mathbf{\widehat{F}}_{\pm}(x,y,z)\exp(-i\omega t)=\\
 &\mathbf{\widehat{F}}_{\pm}(x\mp \beta_x/|\p|,y\mp \beta_y/|\p| ,z\mp \beta_z/|\p|)\exp(-i\omega t),
\end{split}
\end{equation}
where the anticipated spatial translation is explicitly seen in the displacements of the cartesian coordinates.

\begin{figure}[t]
\begin{center}
\includegraphics[scale=0.8]{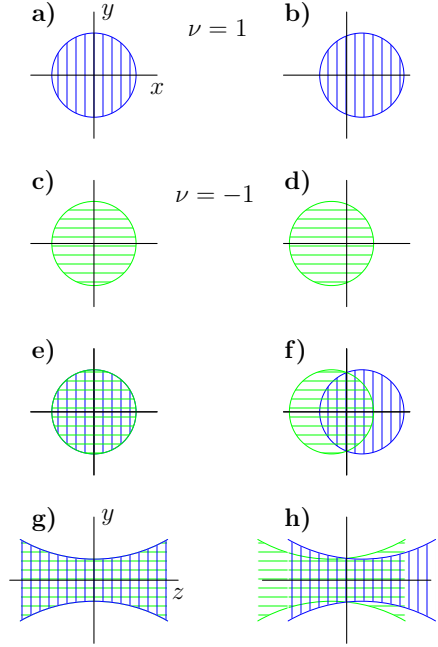}
\end{center}
\caption{(Color online) The diagrams on the left represent the transverse (a,c,e) and longitudinal (g) intensity patterns of Gaussian-like monochromatic fields with different helicity content. The diagrams on the right show the effect that the application of transformations generated by $\Shat$ have on these fields. (a-f) Effect of $\exp(-i\beta_x \hat{S}_x)$ on the transverse intensity pattern for (a, b) a field of well defined helicity equal to one, (c, d) a field of well defined helicity equal to minus one, and (e, f) a field containing both helicity components. (g, f) show the effect of $\exp(-i\beta_z \hat{S}_z)$ on the longitudinal intensity pattern of a field containing both helicity components.}
\label{fig:gauss}
\end{figure}

Equations (\ref{eq:deriv}), (\ref{eq:trans}) and (\ref{eq:F}), provide physical insight into the transformations generated by $\Shat$. Fig. \ref{fig:gauss} depicts the helicity dependent displacement experienced by a monochromatic Gaussian-like field upon application of $\exp(-i\beta_x \hat{S}_x)$ or $\exp(-i\beta_z \hat{S}_z)$. The opposite displacement of the two helicity components of the beam seen in Fig. \ref{fig:gauss} e)-f) and g)-h) illustrates the action of the transformations generated by $\Shat$ on modes of mixed helicity. Note that any electromagnetic field can always be decomposed into modes of well defined helicity.

We now consider the other part of the split: The ``orbital angular momentum'' operator $\Lhat$. Since $\J=\Lhat+\Shat$:
\begin{equation}
	\Lhat=\J-\Shat=\J-\Lambda\frac{\PP}{|\PP|}.
\end{equation}
Since rotations and translations along the same axis commute and helicity commutes with all rotations and translations, the transformations generated by $\Lhat$ are trivially separated into those generated by $\Shat$ and those generated by $\J$. Referring again to the example of monochromatic fields, each component of $\Lhat$, $\hat{L}_i$, produces a helicity dependent translation along the $i$-axis followed by a rotation around the same axis. The order in which the two operations are applied does not matter.

\section{Examples of the use of $\Shat$ and $\Lhat$ in light matter interactions}\label{sec:examples}
With the insight gained up to this point, we can now use $\Shat$ and $\Lhat$ to make some qualitative considerations about light matter interactions. In the following examples, we assume that the frequency of the field is preserved by the interaction with the material system.

An aplanatic lens preserves helicity and the component of angular momentum along its axis \cite[Sec. V B]{FerCor2012b}, say $J_z$, but it does not preserve either $\hat{S}_z$ or $\hat{L}_z$ because the lensing action changes $P_z$. The lens is thus a cylindrically symmetric system that does not preserve either $\hat{S}_z$ or $\hat{L}_z$. On the other hand, the natural modes of a straight electromagnetic waveguide of arbitrary cross-section will be eigenstates of $P_z$, and, if all the materials have the same ratio of electric and magnetic constants, they will be eigenstates of helicity as well \cite{FerCor2012p}. Therefore, $\hat{S}_z$ can be used to classify the eigenmodes of a non cylindrically symmetric system. These two examples illustrate the fact that $\Shat$ and $\Lhat$ are not related to rotations.

We now discuss an application of one kind of simultaneous eigenstates of $\Shat$. Since the three components of $\Shat$ commute, there exist electromagnetic modes of light with simultaneously well defined values for the three of them. We need the eigenvalue of one more independent commuting operator to completely specify an electromagnetic field. If we choose helicity, what we obtain is a plane wave of well defined helicity $|\Psi_{\p,\nu}\rangle$. If we choose parity, which commutes with $\Shat$ since it simultaneously flips the sign of both helicity and momentum, we obtain a so called standing or stationary wave. 
\begin{equation}
	\frac{1}{\sqrt{2}}\left(|\Psi_{\p,+}\rangle \pm |\Psi_{-\p,-}\rangle\right).
\end{equation}
In the coordinate representation, the electric field of such a mode reads, for $\p=p \mathbf{\hat{z}}$: 
\begin{equation}
\label{eq:standing}
\left(\xhat+i\yhat\right)\begin{bmatrix}\cos(pz)\\i\sin(pz)\end{bmatrix}\exp\left(-i\omega t\right),
\end{equation}
where the cosine results from the $+$ sign and the sine from the $-$ sign.

Fields similar to those in (\ref{eq:standing}) were recently predicted to achieve an enhanced interaction with chiral molecules \cite{Tang2010}. This points towards a role for the simultaneous eigenstates of $\Shat$ and parity in the study of the interactions of light with chiral molecules. Incidentally, we note that the name ``superchiral fields'' used in that work can be misleading because of the fact that, while chiral objects fundamentally change under parity, the fields in (\ref{eq:standing}) are eigenstates of parity, and therefore stay invariant after a parity transformation.

\section{Polarization and angular momentum}\label{sec:polarization}
We will now study a common interpretation of the split of the total angular momentum. In the literature, the polarization of a field is often considered to be a contributor to its angular momentum \cite{Oneil2002,Padgett2011,Cameron2012}. We now show that, in general, the polarization degrees of freedom of of the field are completely decoupled from its angular momentum.

We start by identifying what we mean by polarization.

Electromagnetic fields are not scalar objects. They have non-scalar degrees of freedom that we refer to as polarization. The difference between scalar and non-scalar degrees of freedom is very apparent in the procedure for building monochromatic solutions to the vectorial Helmholtz equation \cite[chap. 13.1]{Morse1953}, \cite[chap. VII]{Stratton1941}. For suitable coordinate systems, each linearly independent solution of the scalar Helmholtz equation produces two vectorial transverse solutions. These two solutions are orthogonal to each other and to the ones derived from a different linearly independent scalar solution. One of the two solutions is transverse electric (TE) and the other one is transverse magnetic (TM). Their sum and subtraction produce modes of well defined helicity $(\pm$) \cite[App. A]{FerCor2012b}. Therefore, either TE/TM or helicity ($\pm$) can be used as the label for the polarization, i.e, the degree of freedom that the scalar solution does not have.

On the other hand, the properties of the originating scalar solution determine other properties of the vectorial modes. For example, the scalar function determines the linear momentum of plane waves, one component of angular momentum and the squared angular momentum of multipoles and the same component of angular and linear momentum of Bessel beams \cite[chap. VII]{Stratton1941}. It is important to note that these three families of vector modes are complete orthogonal basis for electromagnetic fields in vacuum.

From this discussion alone, it can already be argued that the polarization and the angular momentum of an electromagnetic field are decoupled degrees of freedom. The argument is that angular momentum is determined by the scalar function, which gives rise to two transverse orthogonally polarized fields. Making arbitrary linear combinations of those two modes will maintain the same angular momentum but vary the polarization degree of freedom through its complete range of possible values. In the general case, polarization cannot affect angular momentum, and viceversa. The same is true for the other properties that the vectorial mode inherits from the scalar solution.

We will now give a formal proof of the idea. Consider the following construction:
\begin{eqnarray}\nonumber
	\label{eq:m}
	&&|\Psi_m\rangle=\Longint\exp(im\phi)R_z(\phi)R_y(\theta)\\	
	&&\left(c_+(\omega,\theta)|\Psi_{[0,0,\omega],+}\rangle+c_-(\omega,\theta)|\Psi_{[0,0,\omega],-}\rangle\right).
\end{eqnarray}

The mode in (\ref{eq:m}) is generated by a linear superposition of plane wave modes. Each plane wave is initially built as a linear superposition of two plane waves of well defined helicity ($\pm$) and initial momentum aligned with the positive $z$-axis, $\p=[0,0,|\p|=\omega]$. The complex coefficients of the linear superposition are $c_{\pm}(\omega,\theta)$. The resulting plane wave is then transformed by successive rotations along the $y$ and $z$ axis $R_z(\phi)R_y(\theta)$: Its momentum changes and ends up pointing towards the $(\theta,\phi)$ direction. On the other hand, rotations commute with helicity, so the helicity of the plane waves does not change upon rotation. Moreover, these particular combination of rotations and initial plane waves do not result in any phase term acquired by the rotated plane wave \cite[eq. 9.7-12]{Tung1985}. This means that the complex coefficients $c_{\pm}(\omega,\theta)$ completely determine the polarization degree of freedom of each rotated plane wave. Therefore, the ensemble of $c_{\pm}(\omega,\theta)$ completely determine the polarization of $|\Psi_m\rangle$.

We claim that the mode in (\ref{eq:m}) is a general mode with a well defined $z$ component of angular momentum equal to $m$. This can be seen by applying a rotation $R_z(\beta)|\Psi_m\rangle$ to (\ref{eq:m}), and verifying that the state transforms into itself times a phase factor $\exp(-im\beta)|\Psi_m\rangle$ with the following steps: Using that a rotation is a linear operator, that two successive rotations along the same axis are equivalent to a single rotation by the sum of the two angles and changing the integration variable\footnote{The shift in the integration interval due to this change is irrelevant because the argument inside the integral is $2\pi$-periodic in the integration variable.} $\phi\rightarrow \phi+\beta$.  Now comes the crucial point: $|\Psi_m\rangle$ has an angular momentum equal to $m$ independently of $c_{\pm}(\omega,\theta)$, that is, independently of polarization. 

The argument holds for arbitrarily small non-null values of $\theta$, and it therefore also applies to electromagnetic fields that fall within the paraxial approximation. The case of a single plane wave is different. The values of $m$ are restricted to $\pm 1$ and determined by its helicity. After setting $d\theta \sin\theta c_{\pm}(\omega,\theta)=d\theta b_{\pm}(\omega,0)\delta(\theta-0)$ and using the facts that $R_y(0)=I$ and $R_z(\phi)|\Psi_{[0,0,\omega],\pm}\rangle=\exp(\mp\phi)|\Psi_{[0,0,\omega],\pm}\rangle$, we see that the integral on $\phi$ only gives a non-zero contribution for $m=+1$ or $m=-1$, and that $b_{+}(\omega,0)$ is the only remaining term when $m=1$ while $b_{-}(\omega,0)$ is the only remaining term when $m=-1$. The angular momentum along the axis of the plane wave does determine its helicity, and viceversa.

\section{$\Shat$ and the Pauli-Lubanski four vector}\label{sec:shat}
We now report a connection of $\Shat$ with relativistic field theory. The $\Shat$ operators are related to the spatial part of a well known object in relativistic field theory: The Pauli-Lubanski four-vector $W_\mu$. The length of the Pauli-Lubanski four vector $W_\mu W^{\mu}$ is one of the Poincare invariants used to classify elementary particles \cite[chap. 10.4.3]{Tung1985}. It is known \cite[expr. 6.6.6]{Penrose1986} that for a massless field: $W_\mu=\Lambda P_\mu$. For the space components $W_k$ ($k=1,2,3$), we then have that:
\begin{equation}
	\label{eq:sw}
	W_k=\Lambda P_k=\Lambda P_k\frac{|\PP|}{|\PP|}=\hat{S}_k |\PP|=\hat{S}_k P_0,
\end{equation}
where the third equality follows from the definition in (\ref{eq:shat}) and the fourth from the assumption of positive frequencies which selects the $P_0=|\PP|$ option and discards the $P_0=-|\PP|$ from the massless condition $P_0^2=|\PP|^2$. As far as we know, relationship (\ref{eq:sw}) has not been reported previously. 

We also note that the four-vector operator $(\hat{X},\mathbf{\hat{\Pi}})$ defined in \cite[Eq. (24)]{Bliokh2011b}, with time component equal to the ``chirality'' ($\hat{X}$) and space component equal to the ``chiral momentum'' ($\mathbf{\hat{\Pi}}$), which, in our notation would be $\hat{X}\equiv \Lambda P_0$ and $\mathbf{\hat{\Pi}}\equiv \Shat P_0$, is exactly the Pauli Lubanski four-vector $(\hat{X},\mathbf{\hat{\Pi}})\equiv(\Lambda P_0,\Shat P_0)=W_\mu$.

\section{Conclusion}
In conclusion, we have studied the operators $\Shat$ and $\Lhat$, commonly proposed as spin and orbital ``angular momenta''. We have seen that the components of $\Shat$ generate helicity and frequency dependent translations. For example, for a monochromatic field, the transformation $\exp(-i\boldsymbol\beta\cdot\Shat)$ translates the two helicity components of the field in the two opposite directions of the $\boldsymbol\beta$ axis. The transformations generated by $\Lhat$ are trivially separated into rotations and the transformations generated by $\Shat$. We have given some examples of the use of $\Shat$ and $\Lhat$ in light matter interactions. In particular, simultaneous eigenstates of $\Shat$ and parity describe standing waves, which have recently been specifically considered for the interaction of light with chiral molecules \cite{Tang2010}. Additionally, we have pointed out a connection between $\Shat$ and the spatial part of the Pauli-Lubanski four-vector in electromagnetism. We have also shown that the polarization degrees of freedom of a general electromagnetic field are decoupled from its angular momentum except in the single plane wave case. The contents of this paper clearly show that $\Shat$ and $\Lhat$ are not directly useful in problems related with the rotational properties of the electromagnetic field. We have already shown in the past \cite{FerCor2012b} that some of those problems can be fully understood using $\J$, $\Lambda$ and their associated transformations: Rotations and electromagnetic duality.  

{\bf Acknowledgments}
This work was funded by the Center of Excellence for Engineered Quantum Systems (EQuS). G.M.-T is also funded by the Future Fellowship program (FF).


\end{document}